\documentclass[prd,floatfix,showpacs]{revtex4}
\usepackage{epsfig}

\newcommand{\eq}[1]{Eq.~(\ref{#1})}

\newcommand{\be}{\begin{equation}}
\newcommand{\ee}{\end{equation}}
\newcommand{\bea}{\begin{eqnarray}}
\newcommand{\eea}{\end{eqnarray}}
\newcommand{\ben}{\begin{eqnarray*}}
\newcommand{\een}{\end{eqnarray*}}

\newcommand{\DS}{Dyson--Schwinger }

\newcommand{\ST}{Slavnov--Taylor }

\newcommand{\w}{\omega}
\newcommand{\e}{\varepsilon}
\newcommand{\al}{\alpha}
\newcommand{\ba}{\beta}
\newcommand{\ga}{\gamma}
\newcommand{\G}{\Gamma}
\newcommand{\de}{\delta}

\newcommand{\si}{\sigma}
\newcommand{\Si}{\Sigma}
\newcommand{\ro}{\rho}
\newcommand{\la}{\lambda}

\newcommand{\et}{\eta}
\newcommand{\ha}{\frac{1}{2}}
\newcommand{\pd}{\partial}

\renewcommand{\th}{\theta}
\newcommand{\cd}{{\cal D}}
\newcommand{\cs}{{\cal S}}
\renewcommand{\div}{\vec{\nabla}}

\newcommand{\s}[2]{{#1}\!\cdot\!{#2}}
\newcommand{\ov}[1]{\overline{#1}}
\newcommand{\dk}[1]{\,\,\,\raisebox{-0.4ex}{\large $\bar{}$}\!\!d\,{#1}\,}
\newcommand{\dx}[1]{d^4{#1}\,}

\newcommand{\ev}[1]{<\!\!{#1}\!\!>}
\begin{document}
\title{Two-Point Functions of Coulomb Gauge Yang-Mills Theory}
\author{P.~Watson}
\author{H.~Reinhardt}
\affiliation{Institut f\"ur Theoretische Physik, Universit\"at T\"ubingen, 
Auf der Morgenstelle 14, D-72076 T\"ubingen, Deutschland}
\begin{abstract}
The functional approach to Coulomb gauge Yang-Mills theory is considered 
within the standard, second order, formalism.  The \DS equations and \ST 
identities concerning the two-point functions are derived explicitly and 
one-loop perturbative results are presented.
\end{abstract}
\maketitle
\section{Introduction}
\setcounter{equation}{0}

Coulomb gauge Yang-Mills theory (and by extension Quantum Chromodynamics) 
is a fascinating, yet frustrating endeavor.  On the one hand, Coulomb 
gauge offers great potential for understanding such issues as confinement 
\cite{Zwanziger:1998ez,Gribov:1977wm}; on the other, the intrinsic 
noncovariance of the formalism makes any perturbative calculation formidably 
complicated.  Many approaches to solving (or providing reliable 
approximations to solving) the problems in Coulomb gauge have been 
forwarded.  Recent among these are the Hamiltonian approach of 
Ref.~\cite{Feuchter:2004mk}, based on the original work of Christ and 
Lee \cite{Christ:1980ku}.  A lattice version of the Coulomb gauge action 
also exists \cite{Zwanziger:1995cv}, which has led to numerical studies, 
for example Refs.~\cite{Cucchieri:2000gu}.  Functional methods based on 
the Lagrangian formalism have also been considered, especially within the 
first order (phase space) formalism \cite{Zwanziger:1998ez,Watson:2006yq} 
and most recently, one-loop perturbative results for both the ultraviolet 
divergent and finite parts of the various two-point functions have been 
obtained \cite{Watson:2007mz}.  Similar results were previously obtained 
for the gluon propagator functions under a different formalism (using the 
chromoelectric field directly as a degree of freedom and without ghosts) 
and using different methods to evaluate the integrals \cite{Andrasi:2003zn}.

In this paper, we consider the (standard, second order) functional approach 
to Coulomb gauge Yang-Mills theory.  We derive the \DS 
equations and \ST identities for the two-point functions that arise in the 
construction and using the techniques of \cite{Watson:2007mz} we present 
results for the one-loop perturbative dressing functions.

The paper is organized as follows.  In the next section, the functional 
formalism used is described.  Section~3 concerns the decomposition of the 
functions used.  The (nonperturbative) \DS equations and \ST identities 
relating the various Green's functions are derived in Section~4.  In 
Section~5, the one-loop perturbative results are obtained.  Finally, there 
is a summary and outlook.

\section{Functional Formalism}
\setcounter{equation}{0}
Let us begin by considering Coulomb gauge Yang-Mills theory.  We use the 
framework of functional methods to derive the basic equations that will 
later give rise to the \DS equations, \ST identities, Feynman rules etc.  
Throughout this work, we will use the notation and conventions established 
in \cite{Watson:2006yq,Watson:2007mz}.  We work in Minkowski space (until 
the perturbative integrals are to be explicitly evaluated) with metric 
$g_{\mu\nu}=\mathrm{diag}(1,-\vec{1})$.  Greek letters 
($\mu$, $\nu$, $\ldots$) denote Lorentz indices, roman subscripts 
($i$, $j$, $\ldots$) denote spatial indices and superscripts 
($a$, $b$, $\ldots$) denote color indices.  We will sometimes also write 
configuration space coordinates ($x$, $y$, $\ldots$) as subscripts where no 
confusion arises.

The Yang-Mills action is defined as
\be
\cs_{YM}=\int\dx{x}\left[-\frac{1}{4}F_{\mu\nu}^aF^{a\mu\nu}\right]
\ee
where the (antisymmetric) field strength tensor $F$ is given in terms of 
the gauge field $A_{\mu}^a$:
\be
F_{\mu\nu}^a
=\pd_{\mu}A_{\nu}^a-\pd_{\nu}A_{\mu}^a+gf^{abc}A_{\mu}^bA_{\nu}^c.
\ee
In the above, the $f^{abc}$ are the structure constants of the $SU(N_c)$ 
group whose generators obey $\left[T^a,T^b\right]=\imath f^{abc}T^c$.  The 
Yang-Mills action is invariant under a local $SU(N_c)$ gauge transform 
characterized by the parameter $\th_x^a$:
\be
U_x=\exp{\left\{-\imath\th_x^aT^a\right\}}.
\ee
The field strength tensor can be expressed in terms of the chromoelectric 
and chromomagnetic fields ($\si=A^0$)
\be
\vec{E}^a=-\pd^0\vec{A}^a-\vec{\nabla}\si^a+gf^{abc}\vec{A}^b\si^c,\;\;\;\;
B_i^a=\epsilon_{ijk}\left[\nabla_jA_k^a-\ha gf^{abc}A_j^bA_k^c\right]
\ee
such that $\cs_{YM}=\int(E^2-B^2)/2$.  The electric and magnetic terms in 
the action do not mix under the gauge transform which for the gauge fields 
is written
\be
A_\mu\rightarrow A'_\mu
=U_xA_\mu U_x^\dag-\frac{\imath}{g}(\pd_\mu U_x)U_x^\dag.
\ee
Given an infinitesimal transform $U_x=1-\imath\th_x^aT^a$, the variation of 
the gauge field is
\be
\de A_{\mu}^a=-\frac{1}{g}\hat{D}_{\mu}^{ac}\th^c
\ee
where the covariant derivative in the adjoint representation is given by
\be
\hat{D}_{\mu}^{ac}=\de^{ac}\pd_{\mu}+gf^{abc}A_{\mu}^b.
\ee

Consider the functional integral
\be
Z=\int\cd\Phi\exp{\left\{\imath\cs_{YM}\right\}}
\ee
where $\Phi$ denotes the collection of all fields.  Since the action is 
invariant under gauge transformations, $Z$ is divergent by virtue of the 
integration over the gauge group.  To overcome this problem we use the 
Faddeev-Popov technique and 
introduce a gauge-fixing term along with an associated ghost term 
\cite{IZ}.  Using a Lagrange multiplier field to implement the gauge-fixing, 
in Coulomb gauge ($\s{\div}{\vec{A}}=0$) we can then write
\be
Z=\int\cd\Phi\exp{\left\{\imath\cs_{YM}+\imath\cs_{fp}\right\}},\;\;\;\;
\cs_{fp}=\int d^4x\left[-\la^a\s{\vec{\nabla}}{\vec{A}^a}
-\ov{c}^a\s{\vec{\nabla}}{\vec{D}^{ab}}c^b\right].
\ee
The new term in the action is invariant under the standard BRS transform 
whereby the infinitesimal gauge parameter $\th^a$ is factorized into two 
Grassmann-valued components $\th^a=c^a\de\la$ where $\de\la$ is the 
infinitesimal variation (not to be confused with the colored Lagrange 
multiplier field $\la^a$).  The BRS transform of the new fields reads
\bea
\de\ov{c}^a&=&\frac{1}{g}\la^a\de\la\nonumber\\
\de c^a&=&-\ha f^{abc}c^bc^c\de\la\nonumber\\
\de\la^a&=&0.
\eea

It is at this point that this work diverges from Ref.~\cite{Watson:2006yq} 
in that we remain here within the standard (second order) formalism.  By 
including source terms to $Z$, we construct the generating functional, 
$Z[J]$:
\be
Z[J]=
\int\cd\Phi\exp{\left\{\imath\cs_{YM}+\imath\cs_{fp}+\imath\cs_s\right\}}
\ee
where
\be
\cs_s=\int d^4x\left[\ro^a\si^a+\s{\vec{J}^a}{\vec{A}^a}+\ov{c}^a\et^a
+\ov{\et}^ac^a+\xi^a\la^a\right].
\ee
It is convenient to introduce a compact notation for the sources and fields 
and we denote a generic field $\Phi_\al$ with source $J_\al$ such that the 
index $\al$ stands for all attributes of the field in question (including 
its type) such that we can write
\be
\cs_s=J_\al\Phi_\al
\ee
where summation over all discrete indices and integration over all 
continuous arguments is implicitly understood.  Expanding the various terms 
we have explicitly
\bea
\cs_{YM}&=&
\int d^4x\left\{-\ha A_i^f\left[\de_{ij}\pd_0^2-\de_{ij}\nabla^2
+\nabla_i\nabla_j\right]A_j^f-A_i^f\pd_0\nabla_i\si^f
-\ha\si^f\nabla^2\si^f
\right.\nonumber\\&&\left.
+gf^{fbc}\left[-(\pd_0A_i^f)A_i^b\si^c-(\nabla_i\si^f)A_i^b\si^c
+(\nabla_jA_k^f)A_j^bA_k^c\right]
+g^2f^{fbc}f^{fde}\left[\ha A_i^b\si^cA_i^d\si^e
-\frac{1}{4}A_i^bA_j^cA_i^dA_j^e\right]\right\}.
\nonumber\\
\eea

The field equations of motion are derived from the observation that the 
integral of a total derivative vanishes, up to boundary terms. The boundary 
terms vanish, although this is not trivial in the light of the Gribov 
problem \cite{Gribov:1977wm} (the reader is directed to 
Ref.~\cite{Watson:2006yq} and references therein for a discussion of this 
topic).  Writing $\cs=\cs_{YM}+\cs_{fp}$, we have that
\be
0=\int\cd\Phi\frac{\de}{\de\imath\Phi_\al}
\exp{\left\{\imath\cs+\imath\cs_s\right\}}.
\label{eq:eom0}
\ee
The explicit form of the field equations of motion is given in 
Appendix~\ref{app:eom}.

In addition to the field equations of motion, there exist identities derived 
by considering the BRS invariance of the action (these eventually form the 
\ST identities).  The BRS transform is continuous and we can regard it as a 
change of variables in the functional integral.  Given that the Jacobian of 
such a change of variables is trivial and that the action is invariant, we 
have that
\bea
0&=&\int\cd\Phi\frac{\de}{\de\imath\de\la}
\exp{\left\{\imath\cs+\imath\cs_s+\imath\de\cs_s\right\}}_{\de\la=0}
\nonumber\\
&=&\int\cd\Phi\exp{\left\{\imath\cs+\imath\cs_s\right\}}
\int d^4x\left[\frac{1}{g}\ro^a\pd_0c^a+f^{abc}\ro^a\si^bc^c
-\frac{1}{g}J_i^a\nabla_ic^a+f^{abc}J_i^aA_i^bc^c+\frac{1}{g}\la^a\et^a
+\ha f^{abc}\ov{\et}^ac^bc^c\right].
\nonumber\\
\eea

So far, the generating functional, $Z[J]$, generates all Green's functions, 
connected and disconnected.  The generating functional of connected Green's 
functions is $W[J]$ where
\be
Z[J]=e^{W[J]}.
\ee
We define the classical fields to be
\be
\Phi_\al=\frac{1}{Z}\int\cd\Phi\,\Phi_\al\exp{\imath\cs}
=\frac{1}{Z}\frac{\de Z}{\de\imath J_\al}.
\ee
The generating functional of proper Green's functions is the effective 
action, $\G$, which is a function of the classical fields and is defined 
through a Legendre transform of $W$:
\be
\G[\Phi]=W[J]-\imath J_\al\phi_\al.
\ee
We introduce a bracket notation for derivatives of $W$ with respect to 
sources and of $\G$ with respect to classical fields (no confusion arises 
since the two sets of derivatives are never mixed):
\be
\ev{\imath J_\al}=\frac{\de W}{\de\imath J_\al},\;\;\;\;
\ev{\imath\Phi_\al}=\frac{\de\G}{\de\imath\Phi_\al}.
\ee
It is now possible to present the field equations of motion in terms of 
proper functions (the \DS equations are functional derivatives of these 
equations).  Using the results listed in Appendix~\ref{app:eom} we have:
\bea
\ev{\imath A_{ix}^a}&=&
-\left[\de_{ij}\pd_{0x}^2-\de_{ij}\nabla_x^2+\nabla_{ix}\nabla_{jx}\right]
A_{jx}^a-\pd_{0x}\nabla_{ix}\si_x^a+\nabla_{ix}\la_x^a
\nonumber\\
&&+gf^{abc}\int\dx{y}\dx{z}\pd_{0x}\de(y-x)\de(z-x)
\left[\ev{\imath J_{iy}^b\imath\ro_z^c}+A_{iy}^b\si_z^c\right]
\nonumber\\
&&-gf^{fac}\int\dx{y}\dx{z}\de(z-x)\nabla_{ix}\de(y-x)
\left[\ev{\imath\ro_y^f\imath\ro_z^c}
+\ev{\imath\ov{\et}_z^c\imath\et_y^f}+\si_y^f\si_z^c+\ov{c}_y^fc_z^c\right]
\nonumber\\
&&+gf^{abc}\int\dx{y}\dx{z}
\left[\de_{ij}\de(z-x)\nabla_{kx}\de(y-x)
+\de_{jk}\de(y-x)\nabla_{ix}\de(z-x)
-\de_{ki}\nabla_{jx}\de(y-x)\de(z-x)\right]\times
\nonumber\\
&&\left[\ev{\imath J_{jy}^b\imath J_{kz}^c}+A_{jy}^bA_{kz}^c\right]
\nonumber\\
&&+g^2f^{fac}f^{fde}\left[\ev{\imath\ro_x^c\imath J_{ix}^d\imath\ro_x^e}
+\si_x^c\ev{\imath J_{ix}^d\imath\ro_x^e}
+\si_x^e\ev{\imath\ro_x^c\imath J_{ix}^d}
+A_{ix}^d\ev{\imath\ro_x^c\imath\ro_x^e}+\si_x^cA_{ix}^d\si_x^e\right]
\nonumber\\
&&-\frac{1}{4}g^2f^{fbc}f^{fde}\de_{jk}\de_{il}
\left[\de^{gc}\de^{eh}(\de^{ab}\de^{di}+\de^{ad}\de^{bi})
+\de^{bg}\de^{dh}(\de^{ac}\de^{ei}+\de^{ae}\de^{ci})\right]\times
\nonumber\\
&&\left[\ev{\imath J_{jx}^g\imath J_{kx}^h\imath J_{lx}^i}
+A_{jx}^g\ev{\imath J_{kx}^h\imath J_{lx}^i}
+A_{lx}^i\ev{\imath J_{jx}^g\imath J_{kx}^h}
+A_{kx}^h\ev{\imath J_{jx}^g\imath J_{lx}^i}
+A_{jx}^gA_{kx}^hA_{lx}^i\right],
\label{eq:adse0}\\
\ev{\imath\si_x^a}&=&
-\pd_{0x}\nabla_{ix}A_{ix}^a-\nabla_x^2\si_x^a
-gf^{fba}\int\dx{y}\dx{z}\de(z-x)\pd_{0x}\de(y-x)
\left[\ev{\imath J_{iy}^f\imath J_{iz}^b}+A_{iy}^fA_{iz}^b\right]
\nonumber\\
&&+gf^{abc}\int\dx{y}\dx{z}\left[\nabla_{ix}\de(y-x)\de(z-x)
+\de(y-x)\nabla_{ix}\de(z-x)\right]\left[\ev{\imath J_{iy}^b\imath\ro_z^c}
+A_{iy}^b\si_z^c\right]
\nonumber\\
&&+g^2f^{fba}f^{fde}\left[\ev{\imath J_{ix}^b\imath J_{ix}^d\imath\ro_x^e}
+A_{ix}^b\ev{\imath J_{ix}^d\imath\ro_x^e}
+\si_x^e\ev{\imath J_{ix}^b\imath J_{ix}^d}
+A_{ix}^d\ev{\imath J_{ix}^b\imath\ro_x^e}+A_{ix}^bA_{ix}^d\si_x^c\right],
\label{eq:sidse0}\\
\ev{\imath\la_x^a}&=&-\nabla_{ix}A_{ix}^a,
\label{eq:ladse0}\\
\ev{\imath\ov{c}_x^a}&=&-\nabla_x^2c_x^a
+gf^{abc}\int\dx{y}\dx{z}\nabla_{ix}\de(y-x)\de(z-x)
\left[\ev{\imath J_{iy}^b\imath\ov{\et}_z^c}+A_{iy}^bc_z^c\right].
\label{eq:ghdse0}
\eea
It is also useful to express the $\la$ equation of motion in terms of 
connected functions:
\be
\xi_x^a=\nabla_{ix}\ev{\imath J_{ix}^a}.
\label{eq:xidse0}
\ee
The identity stemming from the BRS invariance is also best expressed in 
terms of both connected and proper functions and reads:
\bea
0&=&\int\dx{x}\left\{\frac{1}{g}\et_x^a\ev{\imath\xi_x^a}
+\frac{1}{g}\ro_x^a\pd_{0x}\ev{\imath\ov{\et}_x^a}
+f^{abc}\ro_x^a\left[\ev{\imath\ro_x^b\imath\ov{\et}_x^c}
+\ev{\imath\ro_x^b}\ev{\imath\ov{\et}_x^c}\right]
-\frac{1}{g}\left[\frac{\nabla_{ix}}{(-\nabla_x^2)}J_{ix}^a\right]\et_x^a
\right.\nonumber\\&&\left.
+f^{abc}J_{ix}^at_{ij}(x)\left[\ev{\imath J_{jx}^b\imath\ov{\et}_x^c}
+\ev{\imath J_{jx}^b}\ev{\imath\ov{\et}_x^c}\right]
+\ha f^{abc}\ov{\et}_x^a\left[\ev{\imath\ov{\et}_x^b\imath\ov{\et}_x^c}
+\ev{\imath\ov{\et}_x^b}\ev{\imath\ov{\et}_x^c}\right]\right\},
\label{eq:jstid0}
\\
0&=&\int\dx{x}\left\{-\frac{1}{g}\ev{\imath\ov{c}_x^a}\la_x^a
-\frac{1}{g}\ev{\imath\si_x^a}\pd_{0x}c_x^a
-f^{abc}\ev{\imath\si_x^a}\left[\ev{\imath\ro_x^b\imath\ov{\et}_x^c}
+\si_x^bc_x^c\right]
-\frac{1}{g}\left[\frac{\nabla_{ix}}{(-\nabla_x^2)}
\ev{\imath A_{ix}^a}\right]\ev{\imath\ov{c}_x^a}
\right.\nonumber\\&&\left.
-f^{abc}\ev{\imath A_{ix}^a}t_{ij}(x)
\left[\ev{\imath J_{jx}^b\imath\ov{\et}_x^c}+A_{jx}^bc_x^c\right]
+\ha f^{abc}\ev{\imath c_x^a}\left[\ev{\imath\ov{\et}_x^b\imath\ov{\et}_x^c}
+c_x^bc_x^c\right]\right\},
\label{eq:stid0}
\eea
where we have used the common trick of using the ghost equation of motion in 
order to reexpress one of the interaction terms transversely, with the 
transverse projector in configuration space being 
$t_{ij}(x)=\de_{ij}+\nabla_{ix}\nabla_{jx}/(-\nabla_x^2)$.  This 
manipulation will be useful when we consider the \ST identities 
for the two-point functions later on.

At this stage it is useful to explore some consequences of the above 
equations that lead to exact statements about the Green's functions.  
Introducing our conventions and notation for the Fourier transform, we have 
for a general two-point function (connected or proper) which obeys 
translational invariance:
\bea
\ev{\imath J_{\al}(y)\imath J_\ba(x)}&=&
\ev{\imath J_\al(y-x)\imath J_\ba(0)}
=\int\dk{k}W_{\al\ba}(k)e^{-\imath k\cdot(y-x)},
\nonumber\\
\ev{\imath\Phi_{\al}(y)\imath\Phi_\ba(x)}&=&
\ev{\imath\Phi_\al(y-x)\imath\Phi_\ba(0)}
=\int\dk{k}\G_{\al\ba}(k)e^{-\imath k\cdot(y-x)},
\eea
where $\dk{k}=d^4k/(2\pi)^4$.  Starting with \eq{eq:ladse0}, we have that 
the only non-zero functional derivative is
\be
\ev{\imath A_{jy}^b\imath\la_x^a}=\imath\de^{ba}\nabla_{jx}\de(y-x)
=\de^{ba}\int\dk{k}k_je^{-\imath k\cdot(y-x)}
\ee
and all other proper Green's functions involving derivatives with respect 
to the $\la$-field vanish (even in the presence of sources).  In terms of 
connected Green's functions, \eq{eq:ladse0} becomes \eq{eq:xidse0} and the 
only non-zero functional derivative is
\be
\nabla_{ix}\ev{\imath\xi_y^b\imath J_{ix}^a}=-\imath\de^{ba}\de(y-x).
\ee
Because \eq{eq:xidse0} involves the contraction of a vector quantity, the 
information is less restricted than previously.  However, we can write down 
the following (true once sources have been set to zero such that the tensor 
structure is determined):
\bea
\ev{\imath J_{jy}^b\imath J_{ix}^a}&=&
\int\dk{k}W_{AA}^{ba}(k)t_{ij}(\vec{k})e^{-\imath k\cdot(y-x)},
\nonumber\\
\ev{\imath\xi_y^b\imath J_{ix}^a}&=&
\de^{ba}\int\dk{k}\frac{k_i}{\vec{k}^2}e^{-\imath k\cdot(y-x)},
\nonumber\\
\ev{\imath\ro_y^b\imath J_{ix}^a}&=&0,
\eea
where $t_{ji}(\vec{k})=\de_{ji}-k_jk_i/\vec{k}^2$ is the transverse 
projector in momentum space.  These relations encode the transverse nature 
of the vector gluon field.  Turning to \eq{eq:jstid0}, we recognize that if 
we functionally differentiate with respect to $\imath\et_y^d$, again with 
respect to $\imath\xi_z^e$ and set sources to zero, we get that
\be
\ev{\imath\xi_z^e\imath\xi_y^d}=0.
\ee
In effect, the auxiliary Lagrange multiplier field $\la$ drops out of the 
formalism to be replaced by the transversality conditions, as it is 
supposed to.

\section{Feynman Rules and Decompositions}
\setcounter{equation}{0}
Let us now discuss the Feynman rules and general decompositions of Green's 
functions that will be relevant to this work.  The Feynman rules for the 
propagators can be derived from the field equations of motion (written in 
Appendix~\ref{app:eom}) by neglecting the interaction terms and functionally 
differentiating.  Denoting the tree-level quantities with a superscript 
$(0)$, the corresponding equations read:
\bea
J_{ix}^a&=&
\left[\de_{ij}\pd_{0x}^2-\de_{ij}\nabla_x^2+\nabla_{ix}\nabla_{jx}\right]
\ev{\imath J_{jx}^a}^{(0)}+\pd_{0x}\nabla_{ix}\ev{\imath\ro_x^a}^{(0)}
-\nabla_{ix}\ev{\imath\xi_x^a}^{(0)},
\nonumber\\
\ro_x^a&=&\pd_{0x}\nabla_{ix}\ev{\imath J_{ix}^a}^{(0)}
+\nabla_x^2\ev{\imath\ro_x^a}^{(0)},
\nonumber\\
\et_x^a&=&\nabla_x^2\ev{\imath\ov{\et}_x^a}^{(0)}.
\eea
The tree-level ghost propagator is then
\be
\ev{\imath\ov{\et}_x^a\imath\et_y^b}^{(0)}
=-\imath\de^{ab}\int\dk{k}\frac{1}{\vec{k}^2}e^{-\imath k\cdot(y-x)}
\ee
and we identify the momentum space propagator as
\be
W_c^{(0)ab}(k)=-\de^{ab}\frac{\imath}{\vec{k}^2}.
\ee
The rest of the propagators follow a similar pattern and their momentum 
space forms (without the common color factor $\de^{ab}$) are given in 
Table~\ref{tab:w0}.  Note that it is understood that the denominator 
factors involving both temporal and spatial components implicitly carry 
the relevant Feynman prescription, i.e.,
\be
\frac{1}{\left(k_0^2-\vec{k}^2\right)}
\rightarrow\frac{1}{\left(k_0^2-\vec{k}^2+\imath0_+\right)},
\ee
such that the integration over the temporal component can be analytically 
continued to Euclidean space.  It is also useful to repeat this analysis 
for the proper two-point functions and using the tree-level components of 
Eqs.~(\ref{eq:adse0}), (\ref{eq:sidse0}) and (\ref{eq:ghdse0}) we have
\bea
\ev{\imath A_{ix}^a}^{(0)}&=&-\left[\de_{ij}\pd_{0x}^2-\de_{ij}\nabla_x^2
+\nabla_{ix}\nabla_{jx}\right]A_{jx}^a-\pd_{0x}\nabla_{ix}\si_x^a
+\nabla_{ix}\la_x^a,
\nonumber\\
\ev{\imath\si_x^a}^{(0)}&=&-\pd_{0x}\nabla_{ix}A_{ix}^a-\nabla_x^2\si_x^a,
\nonumber\\
\ev{\imath\ov{c}_x^a}^{(0)}&=&-\nabla_x^2c_x^a.
\eea
The ghost proper two-point function in momentum space is
\be
\G_c^{(0)ab}(k)=\de^{ab}\imath\vec{k}^2
\ee
and the rest are presented (without color factors) in Table~\ref{tab:w0}.  
It is immediately apparent that the gluon polarization is \emph{not} 
transverse in contrast to Landau gauge.

\begin{table}
\begin{tabular}{|c|c|c|c|}\hline
$W^{(0)}$&$A_j$&$\si$&$\la$
\\\hline\rule[-2.4ex]{0ex}{5.5ex}
$A_i$&$t_{ij}(\vec{k})\frac{\imath}{(k_0^2-\vec{k}^2)}$&$\underline{0}$&$
\underline{\frac{(-k_i)}{\vec{k}^2}}$
\\\hline\rule[-2.4ex]{0ex}{5.5ex}
$\si$&$\underline{0}$&$\frac{\imath}{\vec{k}^2}$&$\frac{(-k^0)}{\vec{k}^2}$
\\\hline\rule[-2.4ex]{0ex}{5.5ex}
$\la$&$\underline{\frac{k_j}{\vec{k}^2}}$&$\frac{k^0}{\vec{k}^2}$&$
\underline{0}$
\\\hline
\end{tabular}
\hspace{1cm}
\begin{tabular}{|c|c|c|c|}\hline
$\G^{(0)}$&$A_j$&$\si$&$\la$
\\\hline\rule[-2.4ex]{0ex}{5.5ex}
$A_i$&$-\imath k_0^2\de_{ij}+\imath\vec{k}^2t_{ij}(\vec{k})$&
$\imath k^0k_i$&$\underline{k_i}$
\\\hline\rule[-2.4ex]{0ex}{5.5ex}
$\si$&$\imath k^0k_j$&$-\imath\vec{k}^2$&$\underline{0}$
\\\hline\rule[-2.4ex]{0ex}{5.5ex}
$\la$&$\underline{-k_j}$&$\underline{0}$&$\underline{0}$
\\\hline
\end{tabular}
\caption{\label{tab:w0}Tree-level propagators [left] and two-point proper 
functions [right] (without color factors) in momentum space.  Underlined 
entries denote exact results.}
\end{table}

The tree-level vertices are determined by taking the various interaction 
terms of Eqs.~(\ref{eq:adse0}-\ref{eq:ghdse0}) and functionally 
differentiating.  Since, in this study, we are interested only in the 
eventual one-loop perturbative results we omit the tree-level four-point 
functions ($\G_{4A}$ and $\G_{AA\si\si}$).  Defining all momenta as 
incoming, we have:
\bea
\G_{\si AAjk}^{(0)abc}(p_a,p_b,p_c)&=&\imath gf^{abc}\de_{jk}(p_b^0-p_c^0),
\nonumber\\
\G_{\si A\si j}^{(0)abc}(p_a,p_b,p_c)&=&-\imath gf^{abc}(p_a-p_c)_j,
\nonumber\\
\G_{3A ijk}^{(0)abc}(p_a,p_b,p_c)&=&
-\imath gf^{abc}
\left[\de_{ij}(p_a-p_b)_k+\de_{jk}(p_b-p_c)_i+\de_{ki}(p_c-p_a)_j\right],
\nonumber\\
\G_{\ov{c}cA i}^{(0)abc}(p_{\ov{c}},p_c,p_A)&=&-\imath gf^{abc}p_{\ov{c}i}.
\eea

In addition to the tree-level expressions for the various two-point 
functions (connected and proper) it is necessary to consider their general 
nonperturbative structures.  These structures are determined by considering 
the properties of the fields under the discrete transforms of time-reversal 
and parity (the noncovariant analogue of Lorentz invariance arguments for 
covariant gauges).  Using the same techniques as in 
Ref.~\cite{Watson:2006yq} we can easily write down the results in momentum 
space.  For the ghost, we have
\be
W_c^{ab}(k)=-\de^{ab}\frac{\imath}{\vec{k}^2}D_c(\vec{k}^2),\;\;\;\;
\G_c^{ab}(k)=\de^{ab}\imath\vec{k}^2\G_c(\vec{k}^2)
\ee
and the rest are presented in Table~\ref{tab:decomp}.  With the exception 
of the ghost, all dressing functions are scalar functions of \emph{two} 
independent variables, $k_0^2$ and $\vec{k}^2$.  The ghost dressing 
functions are functions of $\vec{k}^2$ only for exactly the same reasons as 
in the first order formalism \cite{Watson:2006yq}.  At tree-level, all 
dressing functions are unity.

\begin{table}
\begin{tabular}{|c|c|c|c|}\hline
$W$&$A_j$&$\si$&$\la$
\\\hline\rule[-2.4ex]{0ex}{5.5ex}
$A_i$&$t_{ij}(\vec{k})\frac{\imath}{(k_0^2-\vec{k}^2)}D_{AA}$&$0$&
$\frac{(-k_i)}{\vec{k}^2}$
\\\hline\rule[-2.4ex]{0ex}{5.5ex}
$\si$&$0$&$\frac{\imath}{\vec{k}^2}D_{\si\si}$&
$\frac{(-k^0)}{\vec{k}^2}D_{\si\la}$
\\\hline\rule[-2.4ex]{0ex}{5.5ex}
$\la$&$\frac{k_j}{\vec{k}^2}$&$\frac{k^0}{\vec{k}^2}D_{\si\la}$&$0$
\\\hline
\end{tabular}
\hspace{1cm}
\begin{tabular}{|c|c|c|c|}\hline
$\G$&$A_j$&$\si$&$\la$
\\\hline\rule[-2.4ex]{0ex}{5.5ex}
$A_i$&$-\imath(k_0^2-\vec{k}^2)t_{ij}(\vec{k})\G_{AA}
-\imath k_0^2\frac{k_ik_j}{\vec{k}^2}\ov{\G}_{AA}$&
$\imath k^0k_i\G_{A\si}$&$k_i$
\\\hline\rule[-2.4ex]{0ex}{5.5ex}
$\si$&$\imath k^0k_j\G_{A\si}$&$-\imath\vec{k}^2\G_{\si\si}$&$0$
\\\hline\rule[-2.4ex]{0ex}{5.5ex}
$\la$&$-k_j$&$0$&$0$
\\\hline
\end{tabular}
\caption{\label{tab:decomp}General form of propagators [left] and two-point 
proper functions [right] (without color factors) in momentum space.  All 
dressing functions are functions of $k_0^2$ and $\vec{k}^2$.}
\end{table}

The dressing functions for the propagators and two-point proper functions 
are related via the Legendre transform.  The connection follows from
\be
\frac{\de\imath J_\ba}{\de\imath J_\al}=\de_{\al\ba}
=-\imath\frac{\de}{\de\imath J_\al}\ev{\imath\Phi_\ba}
=\frac{\de\Phi_\ga}{\de\imath J_\al}\ev{\imath\Phi_\ga\imath\Phi_\ba}
=\ev{\imath J_\al\imath J_\ga}\ev{\imath\Phi_\ga\imath\Phi_\ba}.
\label{eq:leg}
\ee
(Recall here that there is an implicit summation over all discrete indices 
and integration over continuous variables labeled by $\ga$.)  Considering 
all the possibilities in turn, we find that
\bea
D_{AA}&=&\G_{AA}^{-1},\nonumber\\
D_{\si\si}&=&\G_{\si\si}^{-1},\nonumber\\
D_c&=&\G_c^{-1},\nonumber\\
D_{\si\la}&=&\G_{A\si}\G_{\si\si}^{-1}=\ov{\G}_{AA}\G_{A\si}^{-1}.
\label{eq:rnd0}
\eea
Actually, whilst we have included $D_{\si\la}$ up to this point, since 
there is no vertex involving the $\la$-field this propagator will not 
directly play any role in the formalism.  However, indirectly it does turn 
out to have a meaning as will be shown in the next section.

\section{\DS Equations and \ST Identities}
\setcounter{equation}{0}

With the observation that
\be
\frac{\de}{\de\imath\Phi_\ba}\ev{\imath J_\ga\imath J_\al}=
-\ev{\imath J_\ga\imath J_\e}\ev{\imath\Phi_\e\imath\Phi_\ba\imath\Phi_\de}
\ev{\imath J_\de\imath J_\al}
\label{eq:leg1}
\ee
[stemming from the Legendre transform and following from \eq{eq:leg}], the 
derivation of the \DS equations 
becomes relatively straightforward.  Starting with \eq{eq:adse0}, omitting 
the terms that will not contribute at one-loop perturbatively and 
recognizing the tree-level vertices in configuration space, we have that
\bea
\ev{\imath A_{ix}^a}&=&
\imath\left[\de_{ij}\pd_{0x}^2-\de_{ij}\nabla_x^2
+\nabla_{ix}\nabla_{jx}\right]\imath A_{jx}^a
+\imath\pd_{0x}\nabla_{ix}\imath\si_x^a-\imath\nabla_{ix}\imath\la_x^a
\nonumber\\
&&-\int\dx{y}\dx{z}\G_{\si AAij}^{(0)cab}(z,x,y)
\left[\ev{\imath J_{jy}^b\imath\ro_z^c}-\imath A_{jy}^b\imath\si_z^c\right]
-\int\dx{y}\dx{z}\frac{1}{2!}\G_{\si A\si i}^{(0)cab}(z,x,y)
\left[\ev{\imath\ro_y^b\imath\ro_z^c}-\imath\si_y^b\imath\si_z^c\right]
\nonumber\\
&&-\int\dx{y}\dx{z}\frac{1}{2!}\G_{3Aijk}^{(0)abc}(x,y,z)
\left[\ev{\imath J_{jy}^b\imath J_{kz}^c}
-\imath A_{jy}^b\imath A_{kz}^c\right]
+\int\dx{y}\dx{z}\G_{\ov{c}cAi}^{(0)bca}(y,z,x)
\left[\ev{\imath\ov{\et}_z^c\et_y^b}+\imath c_z^c\imath c_y^b\right]
\nonumber\\
&&+\ldots
\eea
Taking the functional derivative with respect to $\imath A_{lw}^f$, 
using \eq{eq:leg1}, setting sources to zero and Fourier transforming to 
momentum space (each step is straightforward so we omit the details for 
clarity) we get the \DS equation for the gluon polarization:
\bea
\G_{AAil}^{af}(k)&=&\de^{af}\left[-\imath(k_0^2-\vec{k}^2)\de_{il}
-\imath k_ik_l\right]\nonumber\\
&&+\int\dk{\w}\G_{\si AAij}^{(0)cab}(\w-k,k,-\w)W_{AAjm}^{bd}(\w)
\G_{\si AAml}^{edf}(k-\w,\w,-k)W_{\si\si}^{ec}(\w-k)
\nonumber\\
&&+\frac{1}{2!}\int\dk{\w}\G_{\si A\si i}^{(0)cab}(\w-k,k,-\w)
W_{\si\si}^{bd}(\w)\G_{\si A\si l}^{dfe}(\w,-k,k-\w)W_{\si\si}^{ec}(\w-k)
\nonumber\\
&&+\frac{1}{2!}\int\dk{\w}\G_{3Aijk}^{(0)abc}(k,-\w,\w-k)W_{AAjm}^{bd}(\w)
\G_{3Amln}^{dfe}(\w,-k,k-\w)W_{AAnk}^{ec}(\w-k)
\nonumber\\
&&-\int\dk{\w}\G_{\ov{c}cAi}^{(0)bca}(\w-k,-\w,k)W_c^{cd}(\w)
\G_{\ov{c}cAl}^{def}(\w,k-\w,-k)W_c^{eb}(\w-k)+\ldots
\label{eq:gldse1}
\eea
Turning now to \eq{eq:sidse0}, we have
\bea
\ev{\imath\si_x^a}&=&\imath\pd_{0x}\nabla_{ix}\imath A_{ix}^a
+\imath\nabla_x^2\imath\si_x^a
-\int\dx{y}\dx{z}\frac{1}{2!}\G_{\si AAjk}^{(0)abc}(x,y,z)
\left[\ev{\imath J_{jy}^b\imath J_{kz}^c}
-\imath A_{jy}^b\imath A_{kz}^c\right]
\nonumber\\
&&-\int\dx{y}\dx{z}\G_{\si A\si j}^{(0)abc}(x,y,z)
\left[\ev{\imath J_{jy}^b\imath\ro_z^c}-\imath A_{jy}^b\imath\si_z^c\right]
+\ldots
\eea
where again, terms that do not contribute at the one-loop perturbative 
level are omitted.  There are two functional derivatives of interest, 
those with respect to $\imath\si_w^f$ and $\imath A_{lw}^f$, which give 
rise to the following two \DS equations:
\bea
\G_{\si\si}^{af}(k)&=&\de^{af}(-\imath\vec{k}^2)+\frac{1}{2!}\int\dk{\w}
\G_{\si AAjk}^{(0)abc}(k,-\w,\w-k)W_{AAjm}^{bd}(\w)
\G_{\si AAmn}^{fde}(-k,\w,k-\w)W_{AAnk}^{ec}(\w-k)
\nonumber\\
&&+\int\dk{\w}\G_{\si A\si j}^{(0)abc}(k,-\w,\w-k)W_{AAjm}^{bd}(\w)
\G_{\si A\si m}^{fde}(-k,\w,k-\w)W_{\si\si}^{ec}(\w-k)+\ldots
\label{eq:sidse1}
\\
\G_{\si Al}^{af}(k)&=&\de^{af}\imath k_0k_l+\frac{1}{2!}\int\dk{\w}
\G_{\si AAjk}^{(0)abc}(k,-\w,\w-k)W_{AAjm}^{bd}(\w)
\G_{3Amln}^{dfe}(\w,-k,k-\w)W_{AAnk}^{ec}(\w-k)
\nonumber\\
&&+\int\dk{\w}\G_{\si A\si j}^{(0)abc}(k,-\w,\w-k)W_{AAjm}^{bd}(\w)
\G_{\si AAml}^{edf}(k-\w,\w,-k)W_{\si\si}^{ec}(\w-k)+\ldots
\label{eq:siadse1}
\eea
Next we consider the ghost equation, \eq{eq:ghdse0}, which can be written
\be
\ev{\imath\ov{c}_x^a}=\imath\nabla_x^2\imath c_x^a+\int\dx{y}\dx{z}
\G_{\ov{c}cAi}^{(0)abc}(x,y,z)\left[\ev{\imath J_{iz}^c\imath\ov{\et}_y^b}
-\imath A_{iz}^c\imath c_y^b\right].
\ee
The ghost \DS equation is subsequently
\be
\G_c^{af}(k)=\de^{af}\imath\vec{k}^2+\int\dk{\w}
\G_{\ov{c}cAi}^{(0)abc}(k,-\w,\w-k)W_c^{bd}(\w)
\G_{\ov{c}cAj}^{dfe}(\w,-k,k-\w)W_{AAji}^{ec}(\w-k).
\label{eq:ghdse1}
\ee

In addition to the \DS equations, the Green's functions are constrained 
by \ST identities.  These are the functional derivatives of \eq{eq:stid0}.  
Since \eq{eq:stid0} is Grassmann-valued, we must first functionally 
differentiate with respect to $\imath c_y^d$.  We are not interested 
(here) in further ghost correlations, so we can then set ghost sources 
to zero.  Also, there is no further information to be gained by considering 
the Lagrange multiplier field $\la^a$, and we set its source to zero also.  
Equation~(\ref{eq:stid0}) then becomes
\bea
\lefteqn{\frac{\imath}{g}\pd_{0y}\ev{\imath\si_y^d}
-f^{abd}\ev{\imath\si_y^a}\imath\si_y^b
-f^{abd}\imath A_{jy}^bt_{ji}(y)\ev{\imath A_{iy}^a}}
\nonumber\\
&=&\int\dx{x}\left\{-f^{abc}\ev{\imath\si_x^a}\frac{\de}{\de\imath c_y^d}
\ev{\imath\ro_x^b\imath\ov{\et}_x^c}+\frac{1}{g}
\left[\frac{\nabla_{ix}}{(-\nabla_x^2)}\ev{\imath A_{ix}^a}\right]
\ev{\imath\ov{c}_x^a\imath c_y^d}
-f^{abc}\ev{\imath A_{ix}^a}t_{ij}(x)\frac{\de}{\de\imath c_y^d}
\ev{\imath J_{jx}^b\imath\ov{\et}_x^c}\right\}.\nonumber\\
\label{eq:stid1}
\eea
Taking the functional derivatives of this with respect to $\imath\si_z^e$ 
or $\imath A_{kz}^e$ and setting all remaining sources to zero gives rise 
to the following two equations:
\bea
\frac{\imath}{g}\pd_{0y}\ev{\imath\si_z^e\imath\si_y^d}
&=&\int\dx{x}\left\{\frac{1}{g}\left[\frac{\nabla_{ix}}{(-\nabla_x^2)}
\ev{\imath\si_z^e\imath A_{ix}^a}\right]\ev{\imath\ov{c}_x^a\imath c_y^d}
\right.\nonumber\\&&\left.
-f^{abc}\ev{\imath\si_z^e\imath\si_x^a}\frac{\de}{\de\imath c_y^d}
\ev{\imath\ro_x^b\imath\ov{\et}_x^c}
-f^{abc}\ev{\imath\si_z^e\imath A_{ix}^a}t_{ij}(x)
\frac{\de}{\de\imath c_y^d}\ev{\imath J_{jx}^b\imath\ov{\et}_x^c}\right\},
\label{eq:stids1}\\
\frac{\imath}{g}\pd_{0y}\ev{\imath A_{kz}^e\imath\si_y^d}
&=&\int\dx{x}\left\{\frac{1}{g}\left[\frac{\nabla_{ix}}{(-\nabla_x^2)}
\ev{\imath A_{kz}^e\imath A_{ix}^a}\right]\ev{\imath\ov{c}_x^a\imath c_y^d}
\right.\nonumber\\&&\left.
-f^{abc}\ev{\imath A_{kz}^e\imath\si_x^a}\frac{\de}{\de\imath c_y^d}
\ev{\imath\ro_x^b\imath\ov{\et}_x^c}
-f^{abc}\ev{\imath A_{kz}^e\imath A_{ix}^a}t_{ij}(x)
\frac{\de}{\de\imath c_y^d}\ev{\imath J_{jx}^b\imath\ov{\et}_x^c}\right\}.
\label{eq:stida1}
\eea
Now, using \eq{eq:leg1}, we have that
\be
f^{abc}\frac{\de}{\de\imath c_y^d}\ev{\imath\ro_x^b\imath\ov{\et}_x^c}=
-f^{abc}\ev{\imath\ov{\et}_x^c\imath\et_\al}
\ev{\imath\ov{c}_\al\imath c_y^d\imath\Phi_\ga}
\ev{\imath J_\ga\imath\ro_x^b}=\de^{ad}\tilde{\Si}_{\si;\ov{c}c}(x,y).
\ee
Taking the Fourier transform
\be
\tilde{\Si}_{\si;\ov{c}c}(x,y)=\int\dk{k}\tilde{\Si}_{\si;\ov{c}c}(k)
e^{-\imath k\cdot(x-y)}
\ee
we get that
\be
\tilde{\Si}_{\si;\ov{c}c}(k)=N_c\int\dk{\w}W_c(k-\w)
\G_{\ov{c}c\ga}(k-\w,-k,\w)W_{\ga\si}(\w).
\ee
Since the ghost Green's functions are independent of the ghost line's 
energy scale \cite{Watson:2006yq}, after $\w_0$ has been integrated out, 
there is no external energy scale and
\be
\tilde{\Si}_{\si;\ov{c}c}(k)=\tilde{\Si}_{\si;\ov{c}c}(\vec{k}).
\label{eq:sicc0}
\ee
However, under time-reversal the $\si$-field changes sign (such that the 
action remains invariant) which in momentum space means that under the 
transform $k_0\rightarrow-k_0$, $\tilde{\Si}_{\si;\ov{c}c}(k)$ must change 
sign and so, given \eq{eq:sicc0} we have the result that
\be
\tilde{\Si}_{\si;\ov{c}c}(k)=0.
\ee
In the case of the term
\be
\de^{af}\tilde{\Si}_{Aj;\ov{c}c}(x,y)=
f^{abc}\frac{\de}{\de\imath c_y^d}\ev{\imath J_{jx}^b\imath\ov{\et}_x^c}
\ee
we can see automatically that in momentum space, 
$\tilde{\Si}_{Aj;\ov{c}c}(k)\sim k_j$ and that the transverse projector 
that acts on it in Eqs.~(\ref{eq:stids1}) and (\ref{eq:stida1}) will kill 
the term.  We thus have
\bea
\frac{\imath}{g}\pd_{0y}\ev{\imath\si_z^e\imath\si_y^d}
&=&\int\dx{x}\left\{\frac{1}{g}\left[\frac{\nabla_{ix}}{(-\nabla_x^2)}
\ev{\imath\si_z^e\imath A_{ix}^a}\right]
\ev{\imath\ov{c}_x^a\imath c_y^d}\right\},
\label{eq:stids2}\\
\frac{\imath}{g}\pd_{0y}\ev{\imath A_{kz}^e\imath\si_y^d}
&=&\int\dx{x}\left\{\frac{1}{g}\left[\frac{\nabla_{ix}}{(-\nabla_x^2)}
\ev{\imath A_{kz}^e\imath A_{ix}^a}\right]
\ev{\imath\ov{c}_x^a\imath c_y^d}\right\},
\label{eq:stida2}
\eea
which in terms of the momentum space dressing functions gives
\bea
\G_{\si\si}(k_0^2,\vec{k}^2)&=&\G_{A\si}(k_0^2,\vec{k}^2)\G_c(\vec{k}^2),
\label{eq:stids3}\\
\G_{A\si}(k_0^2,\vec{k}^2)&=&\ov{\G}_{AA}(k_0^2,\vec{k}^2)\G_c(\vec{k}^2).
\label{eq:stida3}
\eea
The \ST identities for the two-point functions above are rather 
revealing.  They are the Coulomb gauge equivalent of the standard 
covariant gauge result that the longitudinal part of the gluon 
polarization remains bare \cite{Slavnov:1972fg}.  We notice 
that they relate the temporal, longitudinal and ghost degrees of freedom 
in a manner reminiscent of the quartet mechanism in the Kugo-Ojima 
confinement criterion \cite{Kugo:1979gm}.  Also, they 
represent Gau\ss' law as applied to the Green's functions.  
Equation~(\ref{eq:stid1}) suggests that proper functions involving the 
temporal $\si$-field can be systematically eliminated and replaced by 
functions involving the vector $\vec{A}$ and ghost fields although whether 
this is desirable remains to be seen.

We can now return to the general decompositions of the two-point 
functions.  We see that as a consequence of either of the two \ST 
identities above, Eqs.~(\ref{eq:stids3}) or (\ref{eq:stida3}), \eq{eq:rnd0} 
reduces to $D_{\si\la}=D_c$, reassuring us that at least the formalism is 
consistent.  We also see that there are only three independent two-point 
dressing functions, whereas (accounting for the tensor structure of the 
gluon polarization) we have five \DS equations.  We will investigate this 
perturbatively in the next section.

\section{One-Loop Perturbation Theory}
\setcounter{equation}{0}

Let us now consider the one-loop perturbative form of the two-point 
dressing functions that are derived from the \DS equations.  So far, all 
quantities are expressed in Minkowski space.  The perturbative integrals 
must however be evaluated in Euclidean space.  The analytic continuation 
to Euclidean space ($k_0\rightarrow\imath k_4$) is straightforward given 
the Feynman prescription for denominator factors.  Henceforth, all dressing 
functions will be written in Euclidean space and are functions of $k_4^2$ 
and $\vec{k}^2$. The Euclidean four momentum squared is 
$k^2=k_4^2+\vec{k}^2$.  We write the perturbative expansion of the 
two-point dressing functions as follows:
\be
\G_{\al\ba}=1+g^2\G_{\al\ba}^{(1)}.
\ee
The loop integrals will be dimensionally regularized with the (Euclidean 
space) integration measure
\be
\dk{\w}=\frac{d\w_4\,d^d\vec{\w}}{(2\pi)^{d+1}}
\ee
(spatial dimension $d=3-2\e$).  The coupling acquires a dimension:
\be
g^2\rightarrow g^2\mu^\e,
\ee
where $\mu$ is the square of some non-vanishing mass scale squared.  This 
factor is included in $\G_{\al\ba}^{(1)}$ such that the new coupling and 
$\G^{(1)}$ are dimensionless.  By inserting the appropriate tree-level 
factors into the \DS equations, extracting the color and tensor algebra we 
get the following integral expressions for the various two-point proper 
dressing functions:
\bea
(d-1)\G_{AA}^{(1)}(k_4^2,\vec{k}^2)&=&
-N_c\int\frac{\mu^\e\dk{\w}(k_4+\w_4)^2}{k^2\w^2(\vec{k}-\vec{\w})^2}
t_{ij}(\vec{\w})t_{ji}(\vec{k})
-N_c\int\frac{\mu^\e\dk{\w}}{k^2\vec{\w}^2(\vec{k}-\vec{\w})^2}
\w_i\w_jt_{ji}(\vec{k})
\nonumber\\&&
-2N_c\int\frac{\mu^\e\dk{\w}}{k^2\w^2(k-\w)^2}
t_{li}(\vec{k})t_{jm}(\vec{\w})t_{nk}(\vec{k}-\vec{\w})
\left[\de_{ij}k_k-\de_{jk}\w_i-\de_{ki}k_j\right]
\left[\de_{ml}k_n-\de_{ln}k_m-\de_{nm}\w_l\right],
\nonumber\\
\label{eq:dseaa0}\\
\ov{\G}_{AA}^{(1)}(k_4^2,\vec{k}^2)&=&
-N_c\int
\frac{\mu^\e\dk{\w}(k_4+\w_4)^2}{k_4^2\vec{k}^2\w^2(\vec{k}-\vec{\w})^2}
k_ik_jt_{ij}(\vec{\w})
-N_c\int\frac{\mu^\e\dk{\w}}{k_4^2\vec{k}^2\vec{\w}^2(\vec{k}-\vec{\w})^2}
\left[\ha\s{\vec{k}}{(2\vec{\w}-\vec{k})}^2
-\s{\vec{k}}{\vec{\w}}\s{\vec{k}}{(\vec{\w}-\vec{k})}\right]
\nonumber\\&&
-\ha N_c\int\frac{\mu^\e\dk{\w}\s{\vec{k}}{(\vec{k}-2\vec{\w})}^2}{k_4^2
\vec{k}^2\w^2(\vec{k}-\vec{\w})^2}t_{ij}(\vec{\w})t_{ji}(\vec{k}-\vec{\w}),
\label{eq:dseovaa0}\\
\G_{\si\si}^{(1)}(k_4^2,\vec{k}^2)&=&
-\ha N_c\int\frac{\mu^\e\dk{\w}(k_4-2\w_4)^2}{\vec{k}^2\w^2(k-\w)^2}
t_{ij}(\vec{\w})t_{ji}(\vec{k}-\vec{\w})
-4N_c\int\frac{\mu^\e\dk{\w}}{\vec{k}^2\w^2(\vec{k}-\vec{\w})^2}
k_ik_jt_{ij}(\vec{\w}),
\\
\G_{A\si}^{(1)}(k_4^2,\vec{k}^2)&=&
\ha N_c\int\frac{\mu^\e\dk{\w}(k_4-2\w_4)}{k_4\vec{k}^2\w^2(k-\w)^2}
\s{\vec{k}}{(\vec{k}-2\vec{\w})}t_{ij}(\vec{\w})t_{ji}(\vec{k}-\vec{\w})
-2N_c\int\frac{\mu^\e\dk{\w}}{\vec{k}^2\w^2(\vec{k}-\vec{\w})^2}
k_ik_jt_{ij}(\vec{\w}),
\\
\G_c^{(1)}(\vec{k}^2)&=&-N_c\int\frac{\mu^\e\dk{\w}}{\vec{k}^2\w^2(\vec{k}
-\vec{\w})^2}k_ik_jt_{ij}(\vec{\w}).
\eea

At this stage, we are in a position to check the two \ST identities for 
the two-point functions.  The first of these, \eq{eq:stids3}, reads at 
one-loop:
\be
\G_{\si\si}^{(1)}-\G_{A\si}^{(1)}-\G_c^{(1)}=0.
\ee
Inserting the integral expressions above and eliminating overall 
constants, the left-hand side reads
\be
\G_{\si\si}^{(1)}-\G_{A\si}^{(1)}-\G_c^{(1)}\sim
-\ha\int\frac{\dk{\w}(k_4-2\w_4)}{k_4\vec{k}^2\w^2(k-\w)^2}\s{k}{(k-2\w)}
t_{ij}(\vec{\w})t_{ji}(\vec{k}-\vec{\w})
-\int\frac{\dk{\w}}{\vec{k}^2\w^2(\vec{k}-\vec{\w})^2}
k_ik_jt_{ij}(\vec{\w}).
\ee
By expanding the transverse projectors and scalar products, it is 
relatively trivial to show that this does indeed vanish.  The second 
identity, \eq{eq:stida3}, reads
\be
\G_{A\si}^{(1)}-\ov{\G}_{AA}^{(1)}-\G_c^{(1)}=0
\ee
and the left-hand side is:
\bea
\G_{A\si}^{(1)}-\ov{\G}_{AA}^{(1)}-\G_c^{(1)}&\sim&
\ha\int\frac{\dk{\w}\,\s{\vec{k}}{(\vec{k}-2\vec{\w})}}{\w^2(k-\w)^2}
\s{k}{(k-2\w)}t_{ij}(\vec{\w})t_{ji}(\vec{k}-\vec{\w})
+\int\frac{\dk{\w}\,(\w_4^2+2k_4\w_4)}{\w^2(\vec{k}-\vec{\w})^2}
k_ik_jt_{ij}(\vec{\w})
\nonumber\\&&
+\int\frac{\dk{\w}}{\vec{\w}^2(\vec{k}-\vec{\w})^2}
\left[\ha\s{\vec{k}}{(2\vec{\w}-\vec{k})}^2-\s{\vec{k}}{\vec{\w}}
\s{\vec{k}}{(\vec{\w}-\vec{k})}\right].
\eea
Again, it is straightforward to show that this vanishes.  Thus, we have 
reproduced the \ST identity results that tell us that there are only three 
independent two-point dressing functions.

The evaluation of the integrals that give $\G_{AA}$, $\G_{\si\si}$ and 
$\G_c$ is far from trivial.  However, using the techniques developed in 
\cite{Watson:2007mz} it is possible.  For brevity, we do not go into the 
details here and simply quote the results.  They are, as $\e\rightarrow0$:
\bea
\G_{AA}^{(1)}(x,y)&=&
\frac{N_c}{(4\pi)^{2-\e}}
\left\{-\left[\frac{1}{\e}-\ga-\ln{\left(\frac{x+y}{\mu}\right)}\right]
+\frac{64}{9}-3z+g(z)\left[\frac{1}{2z}-\frac{14}{3}+\frac{3}{2}z\right]
-\frac{f(z)}{4}\left[\frac{1}{z}-1+11z-3z^2\right]\right\},
\nonumber\\
\G_{\si\si}^{(1)}(x,y)&=&
\frac{N_c}{(4\pi)^{2-\e}}\left\{
-\frac{11}{3}\left[\frac{1}{\e}-\ga-\ln{\left(\frac{x+y}{\mu}\right)}\right]
-\frac{31}{9}+6z+g(z)(1-3z)-f(z)\left[\ha+2z+\frac{3}{2}z^2\right]\right\},
\nonumber\\
\G_{c}^{(1)}(y)&=&\frac{N_c}{(4\pi)^{2-\e}}\left\{
-\frac{4}{3}\left[\frac{1}{\e}-\ga-\ln{\left(\frac{y}{\mu}\right)}\right]
-\frac{28}{9}+\frac{8}{3}\ln{2}\right\},
\eea
where $x=k_4^2$, $y=\vec{k}^2$, $z=x/y$ and we define two functions:
\bea
f(z)&=&4\ln{2}\frac{1}{\sqrt{z}}\arctan{\sqrt{z}}
-\int_0^1\frac{dt}{\sqrt{t}(1+zt)}\ln{(1+zt)},\nonumber\\
g(z)&=&2\ln{2}-\ln{(1+z)}.
\eea
(The integral occurring in $f(z)$ can be explicitly evaluated in terms of 
dilogarithms \cite{Watson:2007mz}.)  Defining a similar notation for the 
perturbative expansion of the propagator functions:
\be
D_{\al\ba}=1+g^2D_{\al\ba}^{(1)}
\ee
we then have, via \eq{eq:rnd0}, the final results:
\be
D_{AA}^{(1)}(x,y)=-\G_{AA}^{(1)}(x,y),\;\;\;\;
D_{\si\si}^{(1)}(x,y)=-\G_{\si\si}^{(1)}(x,y),\;\;\;\;
D_c^{(1)}(y)=-\G_c^{(1)}(y).
\ee

Several comments are in order here.  Firstly, the expressions for 
$\G_{AA}$ and $\ov{\G}_{AA}$, Eqs.~(\ref{eq:dseaa0}) and 
(\ref{eq:dseovaa0}), respectively, contain energy divergent integrals of 
the form
\be
\int\frac{\dk{\w}\,
\left\{1,\w_i,\w_i\w_j\right\}}{\vec{\w}^2(\vec{k}-\vec{\w})^2}.
\ee
These integrals cancel explicitly, though it should be remarked that this 
cancellation is more obvious in the first order formalism 
\cite{Watson:2007mz}.  Secondly, with respect to the temporal variable $x$, 
all the results above are strictly finite 
for Euclidean and spacelike Minkowski momenta -- any singularities occur 
for $z=x/y=-1$ (the light-cone) with branch cuts extending in the timelike 
direction.  This means that the analytic continuation between Euclidean 
and Minkowski space can be justified.  Thirdly, the coefficient of the 
$\e$-pole for $D_{\si\si}$ and 
the combination $D_{AA}D_c^2$ is $11N_c/3(4\pi)^2$ which is minus the value 
of the first coefficient of the $\ba$-function. This confirms that 
$g^2D_{\si\si}$ \cite{Niegawa:2006ey} and $g^2D_{AA}D_c^2$ (the Coulomb 
gauge analogue of the Landau gauge nonperturbative running coupling) are 
renormalization group invariants at this order in perturbation theory.  
Fourthly, the results above for $D_{AA}$, $D_{\si\si}$ and $D_c$ are 
identical to those calculated within the first order formalism 
\cite{Watson:2007mz}.  

\section{Summary and Outlook}
\setcounter{equation}{0}

The two-point functions (connected and proper) of Coulomb gauge Yang-Mills 
theory have been considered within the standard, second order formalism.  
Functional methods have been used to derive the relevant \DS equations and 
\ST identities.  One-loop perturbative results have been presented and the 
\ST identities that concern them verified.

Suffice it to say that it is tautological for the situation in Coulomb 
gauge to be somewhat different from covariant gauges such as Landau gauge.  
The proper $\vec{A}$-$\vec{A}$ two-point function is explicitly not 
transverse, nor does its longitudinal component remain bare beyond 
tree-level.  This longitudinal component can however be written in terms of 
the temporal gluon and ghost two-point functions via the \ST identities.  
Indeed, the \ST identities show that there are only three independent 
two-point dressing functions: the (transverse) spatial gluon propagator 
dressing function ($D_{AA}$), the temporal gluon propagator dressing 
function ($D_{\si\si}$) and the ghost propagator dressing function 
($D_c$).  With the exception of the ghost dressing function, all are 
noncovariantly expressed in terms of two variables: $k_4^2$ (or $k_0^2$ in 
Minkowski space) and $\vec{k}^2$.  Perturbatively it is seen that the 
analytic continuation between Euclidean and Minkowski space (and vice 
versa) is valid and that the \ST identities hold.

There are many further questions to be addressed.  The perturbative 
structure of the vertex functions, the addition of the quark sector and 
the construction of physical scattering matrix elements from noncovariant 
components are all important next steps.  The issue of noncovariant 
renormalization prescriptions must also be understood.  The connection of 
the functional formalism with other approaches such as the Hamiltonian 
formalism \cite{Feuchter:2004mk} and lattice calculations must also be 
established.  Clearly, there is a lot of work yet to be done.

\begin{acknowledgments}
This work has been supported by the Deutsche Forschungsgemeinschaft (DFG) 
under contracts no. DFG-Re856/6-1 and DFG-Re856/6-2.
\end{acknowledgments}

\appendix
\section{\label{app:eom}Explicit Form of the Field Equations of Motion}
\setcounter{equation}{0}
For completeness, we write the explicit form of the various field equations 
of motion represented by \eq{eq:eom0}:
\bea
J_{ix}^aZ[J]&=&\int\cd\Phi\exp{\left\{\imath\cs+\imath\cs_s\right\}}
\left\{\left[\de_{ij}\pd_{0x}^2-\de_{ij}\nabla_x^2
+\nabla_{ix}\nabla_{jx}\right]A_{jx}^a+\pd_{0x}\nabla_{ix}\si_x^a
-\nabla_{ix}\la_x^a+gf^{fac}\left(\nabla_{ix}\ov{c}_x^f\right)c_x^c
\right.\nonumber\\&&
-gf^{fbc}\left[\de^{af}\pd_{0x}A_{ix}^b\si_x^c
-\de^{ab}\si_x^c\nabla_{ix}\si_x^f+\de^{ab}A_{jx}^c\nabla_{ix}A_{jx}^f
+2\de^{ac}A_{jx}^b\nabla_{jx}A_{ix}^f
-\de^{af}A_{ix}^c\nabla_{jx}A_{jx}^b\right]
\nonumber\\&&
-g^2f^{fac}f^{fde}\si_x^cA_{ix}^d\si_x^e
\nonumber\\&&\left.
+\frac{1}{4}g^2f^{fbc}f^{fde}\left[\de^{ab}A_{jx}^cA_{ix}^dA_{jx}^e
+A_{jx}^b\de^{ac}A_{jx}^dA_{ix}^e+A_{ix}^bA_{jx}^c\de^{ad}A_{jx}^e
+A_{jx}^bA_{ix}^cA_{jx}^d\de^{ae}\right]\right\},
\\
\ro_x^aZ[J]&=&\int\cd\Phi\exp{\left\{\imath\cs+\imath\cs_s\right\}}
\left\{\pd_{0x}\nabla_{ix}A_{ix}^a+\nabla_x^2\si_x^a
-g^2f^{fba}f^{fde}A_{ix}^bA_{ix}^d\si_x^e
\right.\nonumber\\&&\left.
-gf^{fbc}\left[-\de^{ac}A_{ix}^b\pd_{0x}A_{ix}^f
-\de^{ac}A_{ix}^b\nabla_{ix}\si_x^f+\de^{af}\nabla_{ix}A_{ix}^b\si_x^c
\right]\right\},
\\
\xi_x^aZ[J]&=&\int\cd\Phi\exp{\left\{\imath\cs+\imath\cs_s\right\}}
\left\{\nabla_{ix}A_{ix}^a\right\},
\\
\et_x^aZ[J]&=&\int\cd\Phi\exp{\left\{\imath\cs+\imath\cs_s\right\}}
\left\{\nabla_x^2c_x^a-gf^{abc}\nabla_{ix}A_{ix}^bc_x^c\right\}.
\eea


\end{document}